\documentclass[aps,twocolumn,
 superscriptaddress]{revtex4-2} % for review and submission
\usepackage[T1]{fontenc}
\usepackage{tocloft}%need for appendix
\usepackage{graphicx} % needed for figures
\usepackage{dcolumn} % needed for some tables
\usepackage{bm} % for math
\usepackage{amssymb} % for math
\usepackage{amsmath}
\usepackage{mathtools}
\usepackage{cancel}
\usepackage{physics}
\usepackage{float}
\usepackage{afterpage}
\usepackage{hyperref}
\hypersetup{
 colorlinks=true,
 linkcolor=blue,
 filecolor=blue,
 citecolor = black, 
 urlcolor=blue,
 }

\usepackage[switch]{lineno}% Enable numbering of text and display math
\usepackage{orcidlink}

\usepackage{xfrac}

\usepackage{xcolor}
\definecolor{amber}{rgb}{1,0.49,0}

\DeclareFontFamily{U}{mathx}{\hyphenchar\font45}
\DeclareFontShape{U}{mathx}{m}{n}{<-> mathx10}{}
\DeclareSymbolFont{mathx}{U}{mathx}{m}{n}
\DeclareMathAccent{\widebar}{0}{mathx}{"73}

\newcommand{\editor}[2]{%
 \expandafter\newcommand\csname #1note\endcsname[1]{%
 \textcolor{#2}{(\textbf{#1:} ##1)}}%
 \expandafter\newcommand\csname #1\endcsname[1]{%
 \textcolor{#2}{##1}}%
 \expandafter\newcommand\csname #1cancel\endcsname[1]{%
 \textcolor{#2}{\sout{##1}}}%
 \expandafter\newcommand\csname #1change\endcsname[2]{%
 \textcolor{#2}{\sout{##1} ##2}}%
 \newenvironment{#1text}{\color{#2}}{\color{black}}
}

\editor{SB}{amber}
\editor{AF}{blue}
\editor{CAVEAT}{red}
\editor{resub}{black}
\renewcommand{\[}{\begin{equation}}
\renewcommand{\]}{\end{equation}}
\renewcommand{\(}{\begin{equation*}}
\renewcommand{\)}{\end{equation*}}

\DeclarePairedDelimiterX\set[1]\lbrace\rbrace{#1}

\newcommand{\angstrom}{\textup{\AA}}

\newcommand{\ha}{\hat{a}} \newcommand{\had}{\hat{a}^\dagger} \newcommand{\hA}{\widehat{A}}

\newcommand{\hx}{\widehat{X}}

\def\ealla#1{\mathrm e^{#1}}
\def\L{\boldsymbol{\mathcal{L}}}
\def\Q{\boldsymbol{\mathcal{Q}}}

\begin{document}

\title{From Green-Kubo to the full Boltzmann kinetic approach \\ to heat transport in crystals and glasses}

\author{Alfredo Fiorentino,\orcidlink{0000-0002-3048-5534}}
\affiliation{SISSA – Scuola Internazionale Superiore di Studi Avanzati, Trieste, Italy}
\author{Stefano Baroni\,\orcidlink{0000-0002-3508-6663}}
\email{baroni@sissa.it}
\affiliation{SISSA – Scuola Internazionale Superiore di Studi Avanzati, Trieste, Italy}
\affiliation{CNR-IOM DEMOCRITOS, SISSA, Trieste, Italy}
\date{\today}
\begin{abstract}
We show that vertex corrections to the quasi-harmonic Green-Kubo theory of heat transport in insulators naturally lead to a generalisation of the expression for the conductivity that could be derived from the linearized Boltzmann equation, when the effects of the full scattering matrix are accounted for. Our results, which are obtained from the Mori-Zwanzig memory-function formalism, provide a fully \emph{ab initio} derivation of the linearized Boltzmann transport equation and establish a connection between two recently proposed unified approaches to heat transport in insulating crystals and glasses.
\end{abstract}
\maketitle

\resub{\section{Introduction}}
The Green-Kubo (GK) theory of linear response \cite{green1952,*green1954,kubo1957a,*kubo1957b} and the Boltzmann's kinetic approach (BKA), which leads to the transport equation with the same name (BTE) \cite{BTE:Ziman2001}, are usually considered as independent and complementary methods to deal with charge and heat transport in condensed matter. By construction, the GK theory only applies to small perturbations, but it is otherwise general, for it applies to liquids, as well as to solids, either crystalline or amorphous, both in the classical and fully quantum regimes. The BTE is not limited to the linear regime, but it is based on a (semi-) classical treatment of charge and energy, which requires a proper definition of their carriers (\emph{e.g.} electron and phonon quasi-particles) with well defined values of their positions, momenta, and energies \cite{Peierls1929}. These requirements make it problematic to apply the BKA to amorphous solids.

A couple of recent papers \cite{Simoncelli2019,Isaeva2019} have recently generalized these approaches, so as to encompass crystalline and disordered systems in the same theoretical and computational frameworks. The two papers differ conceptually in that the first is based on an elaborate generalization of the BKA, based on Wigner's dynamics (WD) \cite{Simoncelli2021}, while the second is a straightforward specialization of the GK theory to solids in the quasi-harmonic (QH) approximation, and dubbed therefore ``QHGK''. Notwithstanding, the two approaches give similar results for the heat conductivity, which coincide in the long-life-time limit. Even when applied to crystals, these methods provide further insight into the limits of the quasi-particle picture of transport: the quasi-particle group velocities are replaced here with a anti-Hermitian matrix whose diagonal elements are indeed group velocities, and whose off-diagonal elements, which fully determine the transport mechanism in disordered systems, give rise to \emph{inter-band} contributions to heat transport in crystals.

The conceptual content of these two approaches and the meaning of the physical approximations leading to them are best appreciated in a many-body framework \cite{Caldarelli2022}, starting from the GK formula \cite{kubo1957a,*kubo1957b}, which states that the heat conductivity is proportional to the integral of the time correlation function of the energy flux, $\langle\widehat{J}(t)\widehat{J}(0)\rangle$, where $\langle\cdot\rangle$ indicates an equilibrium thermal average and here and a caret, ``$\widehat{\phantom{a}}$'', designates quantum mechanical operators. In the (quasi-) harmonic approximation, the energy flux is quadratic in the phonon (or, more generally, normal-mode) creation and annihilation operators, $\ha_p^\dag $ and $\ha_p$. The energy-flux correlation function is therefore a linear combination of products of four normal-mode operators, of the form $\langle \ha_p^\dag(t) \ha_q(t) \ha_r^\dag \ha_s\rangle$ and $\langle \ha_p^\dag(t) \ha_q^\dag(t) \ha_r \ha_s \rangle$ \cite{Isaeva2019}. The QHGK approach essentially results from the application of two related, but distinct, approximations. The first amounts to factorizing four-point correlation functions into linear combinations of products of two-point ones, such as, \emph{e.g.}, 
\begin{multline} 
 \langle \ha_p^\dag(t) \ha_q(t) \ha_r^\dag \ha_s\rangle \approx n_p n_r
 % \langle\ha_p^\dag(t)\ha_p(t)\rangle\langle\ha_r^\dag(0)\ha_r(0)\rangle
 \delta_{pq}\delta_{rs} + \\ \langle\ha_p^\dag(t)\ha_p(0)\rangle\langle\ha_q(t)\ha_q^\dag(0)\rangle\delta_{ps}\delta_{qr}, 
\end{multline}
where $n_p=\langle \ha_p^\dagger \ha_p \rangle=1/(e^{\hbar\omega_p/k_bT}-1)$ is a Bose-Einstein occupation number, $\omega_p$ being the normal-mode frequency, $T$ the system's temperature, and $k_B$ the Boltzmann's constant. In the many-body parlance, this factorization is described as the neglect of vertex corrections to the correlation function and referred to as the \emph{dressed-bubble approximation} \cite{Caldarelli2022}. Physically, vertex corrections describe the correlation between the decay channels of different normal modes, and their neglect amounts to expressing the propagation and decay of each of them independently from all the others, as if determined by the interaction with a common, mean-field-like, heat bath. The second approximation consists in assuming that this heat bath is essentially a white noise, so that its interaction with the normal modes is Markovian, \emph{i.e.} unaffected by any memory effects. The Markovian approximation essentially results in a damped exponential dependence of the single-mode \emph{greater} Green's functions on time,
\begin{align}\label{eq:gMarkovian}
 -i\langle \ha_p(t) \ha_p^\dag \rangle \approx -i(n_p+1) \mathrm{e}^{-i\omega_pt-\gamma_p|t|},
\end{align}
which is to say that its spectral function (the imaginary part of its Laplace-Fourier transform) is Lorentzian. In a many-body framework, this roughly corresponds to neglecting the frequency dependence of the phonon self-energies. The combination of these two approximations leads to a unified approach to heat transport that---while reducing in crystals to an enhanced version of the BKA in the so-called \emph{relaxation-time approximation} (RTA), where inter-band effects are explicitly accounted for---equally applies to amorphous systems as well.

In this paper we show how the two approximations that lead to the QHGK approach can be removed by treating the anharmonic decay of the vibrational normal modes through the Mori-Zwanzig (MZ) memory-function formalism \cite{Zwanzig1960,Mori_1965}. Our main result is that a proper account of vertex corrections in the QHGK approach leads to an expression for the heat conductivity that, while applying to both lattice-periodic and disordered systems, in the former case naturally reduces to the \emph{full} BTE beyond the RTA, \emph{i.e.} properly accounting for the effects of the full scattering matrix. This result shows how the full linearized BTE can be derived entirely from first principles within the GK theory of linear response and sheds light onto the conceptual equivalence of the QHGK \cite{Isaeva2019} and WD-BKA \cite{Simoncelli2019} approaches to heat transport. \resub{In Sec. \ref{sec:GKMZ} we briefly present the Mori-Zwanzing approach to the Green-Kubo linear-response theory of heat transport; in Sec. \ref{sec:SMA} we derive the single-mode approximation to the GKMZ theory, which is closely related to the QHGK approach of Ref. \cite{Isaeva2019}; in Sec. \ref{sec:GKMZ-BTE} we derive the full Boltzmann equation from the GKMZ approach to heat transport; in Sec. \ref{sec:WTE} we draw a comparison with the WTE approach of Refs. \cite{Simoncelli2019,Simoncelli2021}; in Sec. \ref{sec:Li3ClO} we present a numerical application to the Li$_3$ClO anti-perovskite ionic conductor; Sec. \ref{sec:conclusions} finally contains our conclusions.} 

\resub{\section{Green-Kubo-Mori-Zwanzig theory of lattice heat conductivity}} \label{sec:GKMZ}

The quantum GK formula for the thermal conductivity, $\kappa$, reads:
\begin{equation} \label{eq:GK}
 \kappa = \frac{1}{VT} \int_0^\infty dt \int_0^{\frac{1}{k_BT}} d\lambda \langle\widehat{J}(t-i\hbar\lambda)\widehat{J}(0)\rangle,
\end{equation}
where $V$ is the system's volume, $\widehat J$ is a generic Cartesian component the energy-flux operator in the Heisenberg representation, and Cartesian indices have been suppressed to unclutter the notation. All the algebra that follows is considerably simplified by introducing the Kubo inner product \cite{Kubo_1966,Forster_2018} between quantum mechanical operators, defined as:
\begin{equation} \label{eq:scalar_product}
 \bigl (\widehat A,\widehat B\bigr ) \doteq \int_0^\frac{1}{k_BT} \langle \widehat A^\dag(-i\hbar\lambda) \widehat B \rangle d\lambda,
\end{equation}
where the time evolution of operators in the Heisenberg representation can be formally expressed in terms of the exponential of the Liouvillian super-operator \cite{Fano:1957}, $\L$, defined as $\widehat A(t)= \ealla{i\widehat H t/\hbar} \widehat A \ealla{-i\widehat H t/\hbar} \doteq \ealla{i\L t} \widehat A$, and $\L \widehat A \doteq [\widehat H,\widehat A]/\hbar = -i \dot{\widehat A}$. In terms of this scalar product, the quantum heat conductivity, Eq. \eqref{eq:GK}, simply reads $ \kappa = \frac{1}{VT} \int_0^\infty \bigl ( \widehat J(t), \widehat J(0) \bigr ) dt $, in close formal analogy with its classical counterpart.

In the QH approximation, the energy flux can be cast into the form:
\begin{multline} \label{eq:J-QHA}
 \widehat J =-\frac{i\hbar}{2}\sum_{pq}v_{pq}\left(\frac{\omega_p+\omega_q}{2}\right) \bigr (\ha^\dag_p\ha_q- \ha_q^\dag\ha_p \bigl ) \\
 +\frac{i\hbar}{2}\sum_{pq}v_{pq}\frac{\omega_p-\omega_q}{2} \bigl (\ha^\dag_p\ha^\dag_q- \ha_p\ha_q \bigr ),
\end{multline}
where $v_{pq}$ is the real anti-symmetric generalized group-velocity matrix \cite{Isaeva2019,vMatrixNote}. The first, phonon-conserving, term in Eq. \eqref{eq:J-QHA} was dubbed \emph{resonant} in Ref. \cite{Isaeva2019}, while the second, \emph{anti-resonant}, one was shown to give negligible contributions and will thus not be considered any further in the present work. By applying the GK formula, Eq. \eqref{eq:GK}, and manipulating the indices on account of the antisymmetry of the group-velocity matrix---which makes the ordering of equal-time products of bosonic operators such as, \emph{e.g.}, $\had_p(t)\ha_q(t)$---irrelevant, one finally obtains:
% \begin{multline}\label{eq:GK_manybody}
% \kappa^{\alpha\beta}=\frac{\hbar^2}{VT}\sum_{pqrs}v_{pq}^{\alpha}v_{rs}^{\beta}\left(\frac{\omega_p+\omega_q}{2}\right)\left(\frac{\omega_r+\omega_s}{2}\right)\times\\
% \int_0^\infty dt~\int_0^\frac{1}{k_bT}\langle\ha^\dag_p(\tau)\ha_q(\tau)\ha^\dag_s\ha_r\rangle
% \end{multline}
\begin{equation} \label{eq:GK_manybody}% \label{eq:kappa-1}
 \kappa = \frac{\hbar^2}{VT}\sum_{IJ}v_Iv_J\widebar\omega_I\widebar\omega_J \widebar C_{IJ}(0),
\end{equation}
%
% \begin{equation}\label{eq:GK_manybody}
% \kappa=\frac{\hbar^2}{VT}\sum_{IJ}v_{I}v_{J} \widebar\omega_I \widebar\omega_J \int_0^\infty \Bigl ( \hA_I(t)\hA_J\bigr ) dt,
% \end{equation}
where the indices $I=(pq)$ and $J=(rs)$ label pairs of normal modes, $\widebar \omega_I=(\omega_p+\omega_q)/2$, $\hA_I = \had_p\ha_q$, and $\widebar{C}_{IJ}(0)$ is the zero-frequency value of the Fourier-Laplace (FL) transform of the two-mode correlation function:
\begin{align} \label{eq:CIJdef}
 \begin{aligned}
 \widebar{C}_{IJ}(z)&=\int_0^\infty dt \, \ealla{izt} C_{IJ}(t), \text{ and} \\
 C_{IJ}(t) &= \bigl (\widehat A_I(t), \widehat A_J \bigr ).
 \end{aligned}
\end{align}
For the sake of clarity, we stress that velocity matrices appearing in the expression of vector quantities (such as a current) carry a Cartesian index, whereas products of two such matrices appearing in tensor quantities (such as the conductivity) carry two indices. Eq. \eqref{eq:GK_manybody} shows that the computation of the heat conductivity reduces to that of two-phonon correlation functions. In the following, we show how this task can be effectively tackled by leveraging the MZ memory-function formalism.

In his celebrated 1965 paper \cite{Mori_1965}, Mori showed that $\widebar C_{IJ}(z)$ can be formally expressed as:
\begin{align}
 \widebar C_{IJ}(z) &= i\sum_K\widebar \Lambda_{IK}^{-1}(z) C_{KJ}, \text{ where} \label{eq:C_tilde} \\
 C_{IJ} &= C_{IJ}(0), \label{eq:CIJ(0)}
 \end{align}
 \begin{align}
 \widebar \Lambda_{IK}(z) &= z\delta_{IK}-\Omega_{IK}+i\widebar\Gamma_{IK}(z), \label{eq:Lambda} \\
 \Omega_{IJ} &= i\sum_K \bigl ( \dot{\widehat A}_I, \widehat A_K \bigr )C^{-1}_{KJ}, \label{eq:OmegaIJ}\\
 \widebar \Gamma_{IJ}(z) &= \sum_K \Bigl( \dot{\widehat A}_I,\Q (z-\Q\L\Q)^{-1}\Q \dot{\widehat A}_K \Bigr )C_{KJ}^{-1}, \label{eq:Gamma}
\end{align}
and $\Q$ is the projector over the operator manifold orthogonal to $\mathsf{span}\bigl (\{ \widehat A_I \} \bigr )$, defined by its action onto a generic operator, $\widehat B$, as:
\begin{equation}
 \Q\widehat B = \widehat B -\sum_{IJ}\widehat A_I \bigl ( \widehat A_J, \widehat B \bigr ) C^{-1}_{IJ}. 
\end{equation} 
Mind the difference between $C_{IJ}$, which is a function of time, and its FL transform $\widebar C_{IJ}$, which is a function of frequency. More details on the Mori-Zwanzig formalism can be found, \emph{e.g.}, in Chapter 5 of Ref. \cite{Forster_2018}.

\resub{\subsection{The single-mode approximation}} \label{sec:SMA}
In order to proceed further, we seek to evaluate $\widebar C_{IJ}(z)$, Eq. \eqref{eq:C_tilde}, to leading order in the strength of the anharmonic interactions, $\widehat{\mathcal V}=\widehat H-\widehat H^\circ$, $\widehat H^\circ = \sum_p \hbar \omega_p \bigl (\had_p\ha_p+\sfrac{1}{2} \bigr )$ being the harmonic Hamiltonian. Let us start with $C_{IJ}$, Eq. \eqref{eq:CIJ(0)}, and $\Omega_{IJ}$, Eq. \eqref{eq:OmegaIJ}, whose leading order in $\mathcal V$ is $\mathcal O(1)$. In the harmonic approximation, one has:
\begin{align}
 \begin{aligned}
 C^\circ_{IJ} &= \int_0^\frac{1}{k_bT} d\lambda \langle\ha_q^\dag\ha_s\rangle^\circ \langle\ha_p\ha_r^\dag\rangle^\circ \ealla{\hbar\lambda(\omega_p-\omega_q)}\delta_{ps}\delta_{qr} \\
 &=\frac{n_p-n_q}{\hbar(\omega_q-\omega_p)}\delta_{IJ}, \label{eq:CIJ_harm}
 \end{aligned}
\end{align}
where $\langle\cdot\rangle^\circ$ indicates a thermal average in the canonical ensemble of the harmonic system, and
\begin{equation} \label{eq:OmegaDiag}
 \Omega_{IJ} = (\omega_p-\omega_q)\delta_{IJ}.
\end{equation}
In this approximation the time evolution of $\widehat A_I$ is parallel to $\widehat A_I$: one concludes that $\Q \dot{\widehat A}_I \sim \mathcal O(\mathcal V)$, and that $\widebar\Gamma$ is $\sim \mathcal O(\mathcal V^2)$ (Eq. \ref{eq:Gamma}). We notice that, to this order in $\mathcal V$, the thermal averages that implicitly appear in the Kubo inner product in Eq. \eqref{eq:Gamma} can be performed in the canonical ensemble of the harmonic Hamiltonian: we indicate this ``harmonic'' Kubo product by the symbol $\bigl ( \widehat A,\widehat B \bigr )^\circ$. The $\widebar \Lambda$ matrix in Eq. \eqref{eq:Lambda} is singular at $z=0$ for $\widebar\Gamma=0$: its inverse in Eq. \eqref{eq:C_tilde}, is therefore $\widebar \Lambda^{-1}(0) \sim \mathcal O(\mathcal V^{-2})$, consistently with the divergence of the heat conductivity in the harmonic limit. By using Eq. \eqref{eq:CIJ_harm}, the heat conductivity, Eq. \eqref{eq:GK_manybody}, can be cast into the form:
\begin{align}\label{eq:GK_manybody_symm}
 \kappa &= \frac{i\hbar^2}{VT}\sum_{IJ} v_Iv_J\widebar\omega_I\widebar\omega_J \widebar \Lambda^{-1}_{IJ}(0) C^\circ_{JJ} \nonumber \\
 &= \frac{i}{V}\sum_{IJ}c_J v_I v_J \frac{\omega_I}{\omega_J}\widebar \Lambda^{-1}_{IJ}(0),
\end{align}
where $c_J=\hbar^2\widebar\omega_J^2C_{JJ}/T$ is the resonant generalized modal specific heat introduced in Ref. \cite{Isaeva2019}.
% \begin{align}
% \widebar\Lambda_{IJ}(z)&=(z-\Omega_{II})\delta_{IJ}+i\widebar\Gamma_{IJ}(z) \\
% \widebar\Gamma_{IJ}(z) &= \Gamma_{IJ}(z)\sqrt{\frac{C_{JJ}}{C_{II}}}\label{eq:Gamma_tilde},
% \end{align}

%\AF{and we used that $\widebar C_{IJ}=\frac{d_I}{d_J}C_{IJ} \Rightarrow \widebar C_{IJ}^{-1}=\frac{d_I}{d_J}C_{IJ}^{-1}$ }

Eq. \eqref{eq:GK_manybody_symm} is completely general, though its implementation requires the computation and inversion of the $N^2\times N^2$ matrix, $\widebar \Lambda_{IJ}$, where $N$ is the number of normal modes. This task is greatly facilitated when the off-diagonal elements of this matrix, \emph{i.e.} of the $\widebar\Gamma$ matrix in Eqs. \eqref{eq:Lambda} and \eqref{eq:Gamma}, can be neglected. In this case, Eq. \eqref{eq:GK_manybody_symm} can be shown to reduce to:
\begin{equation} \label{eq:QHGK_limit}
 \kappa =\frac{1}{V}
 \sum_{pq}c_{pq} v_{pq} v_{pq} 
 \frac{ \widebar \Gamma_{pq}^{'}}{(\omega_q-\omega_p-\widebar \Gamma_{pq}^{''} )^2+(\widebar \Gamma_{pq}^{'})^2},
\end{equation}
where $\widebar \Gamma'_{pq} = \mathsf{Re}\, \widebar \Gamma_{pq,pq}(0)$ and $\widebar\Gamma''_{pq} = \mathsf{Im}\, \widebar \Gamma_{pq,pq}(0)$, which is closely reminiscent of QHGK expression of Ref. \cite{Isaeva2019}. The details of the derivation are reported 
%in Sec.~S-I of the supplementary material (SM). 
\resub{Appendix \ref{sec:S-MemoryDiagonal}. }

\resub{The neglect of the off-diagonal elements of $\widebar\Gamma$ is conceptually analogous to the single-mode (SM) approximation in the solution of the linearized BTE. This approximation has been shown to be particularly crude in the presence of strong hydrodynamic effects, especially (but not limited to) 2D materials \cite{Cepellotti2015}, leading to an underestimate of the heat conductivity. In Sec. \ref{sec:GKMZ-BTE} we will further elaborate on this analogy and show how the full BTE can be derived from the GKMZ theory. Some numerical evidence on the magnitude of the effects of non-diagonal terms in $\Gamma$ will be provided in Sec. \ref{sec:Li3ClO}.} 

\resub{Yet, no assumptions have been made on the time dependence of the single-mode Green's function or on the frequency dependence of its Fourier transform. Indeed, in Appendix \ref{sec:diag->QHGK} we show that for a cubic anharmonic potential} Eq. \eqref{eq:QHGK_limit} can be put into the \emph{dressed-bubble} form of Refs. \cite{Caldarelli2022,Dangi2021}:
\begin{align}
 \kappa &= \frac{1}{ V}\sum_{pq}\frac{\hbar^2(\omega_p+\omega_q)^2}{4}v_{pq} v_{pq} (I_{pq}+I_{qp}), \label{eq:gk_product_green_f} \\ %\text{ where} \\
 I_{pq}&=\frac{1}{8\pi k_bT^2}\int d\omega \tilde g_p^>(\omega) \tilde g^<_q(\omega), %\text{ where}
 \label{eq:integral_overlap_spectrum} 
\end{align}
where
\begin{align}
 g^>_p(t) &= -i \langle \ha_p(t) \ha_p^\dag \rangle \\
 g^<_p(t) &= \phantom{-}i \langle \ha_p^\dagger \ha_p(t) \rangle
\end{align}
are the so-called \emph{greater} and \emph{lesser} Green's function, and $\Tilde g^{\scriptscriptstyle\lessgtr}_q(\omega)$ indicate their Fourier transforms, \resub{whose lineshape is not assumed \emph{ a priori}.} In Eq. \eqref{eq:gk_product_green_f} we have omitted terms of order $\mathcal O(N^{-1})$, which are negligible in the thermodynamic limit. 

\resub{In the Markovian approximation, \emph{i.e.} by further neglecting memory effects (which in the many-body parlance amounts to neglecting the frequency dependence of the phonon self-energy), we arrive at the RTA approximation for the SM Green's function, reading:} 
% With the further addition the Markovian approximation, which together with the single mode approximation is equivalent to the RTA approximation}, one has: 
$g^>_p(t)=-i(n_p+1)e^{-i\omega_pt -\gamma_p|t|}$ and $g^<_p(t)=i\,n_pe^{-i\omega_pt -\gamma_p|t|}$. 
%
% \SBnote{ Secondo me qua si rischia di far confusione. La SM l'abbiamo introdotta come un'approssimazione all'equazione di Boltzmann: \`e un'approssimazione alle funzioni di Green a due particelle che mi pare equivalente alla nostra ``dressed-bubble approximation''. Invece, la RTA \`e un'approssimazione alla funzione di Green ad una particella che noi altrove chimiamo ``Markoviana''. Se non mi sono incasinato, \`e forse il caso di fare chiarezza? Cio\`e, se non mi sbaglio, noi chiamiamo nel nostro schema``dressed-bubble'' e ``Markovian'' quello che in BTE si chiama, rispettivamente, SM e RT. O no .?... Se non mi sono incasinato, \`e forse il caso di fare chiarezza? }
%
By plugging these expressions into Eq. \eqref{eq:integral_overlap_spectrum}, one obtains:
\begin{align}
 \kappa^M &= \frac{1}{ V}\sum_{pq}c_{pq}^Mv_{pq} v_{pq}\tau_{pq}, %\text{ where} 
 \label{eq:k_markovian}
\end{align}
where
\begin{align}
 c_{pq}^M &= \frac{n_p(n_q+1)+n_q(n_p+1)}{2k_BT^2}\frac{(\omega_p+\omega_q)^2}{4}, %\text{ and} \nonumber 
 \\
 \tau_{pq} &= \frac{\gamma_p+\gamma_q}{(\omega_p-\omega_q)^2+(\gamma_p+\gamma_q)^2}, % \nonumber \\
% &= \frac{1}{ V}\sum_{pq}\frac{n_p(n_q+1)+n_q(n_p+1)}{2k_bT^2}\frac{(\omega_p+\omega_q)^2}{4}v_{pq} v_{pq}\\ &\times\frac{\gamma_p+\gamma_q}{(\omega_p-\omega_q)^2+(\gamma_p+\gamma_q)^2},
\end{align}
and ``$M$'' stands for ``Markovian'', which coincides with the QHGK expression of Ref. \cite{Isaeva2019}, to within terms of order $\mathcal O(\gamma^2)$, in agreement with the conclusions of Ref. \cite{Caldarelli2022}. \resub{Memory effects, defined in the MZ formalism as the difference between the exact Green's function and the Markovian ansatz, have been shown to have a non-negligible impact on the lattice thermal conductivity of a ferro-electric material near the critical temperature \cite{Dangi2021}.}
%\resub{ The Markovian approximation is largely used and tested to describe the phonon Green's function, however a recent work by \emph{Dangic et al.} \cite{Dangi2021} showed that in a material near ferro-electric transition there are corrections to the lattice thermal conductivity if the true Green's function is used.}

\resub{\section{From GKMZ to the full BTE} \label{sec:GKMZ-BTE}}
Until now, no assumptions on the crystalline order of the system have been made. Lattice periodicity brings about a great simplification that allows one to reduce the MZ approach to heat transport to the \emph{full} (\emph{i.e.} beyond the RTA) BTE. The crucial step permitting this reduction is the realization that lattice periodicity implies a Bloch block structure (no pun intended!) of the velocity matrices. In a periodic system, normal-mode indices split into a pair of Bloch-wavevector and phonon-band indices: $q\to \{ \bm q, \nu \}$ and the velocity matrices read:
\begin{equation}
 \begin{aligned}
 v_{\bm q\nu,\bm k\mu} = v_{\bm q,\nu\mu}\, \delta_{\bm{qk}}, \text{ with} \\
 v_{\bm q,\nu\mu}=v_{\bm q,\mu\nu}^*=-v_{-\bm q,\mu\nu},
 \end{aligned}
\end{equation}
where the diagonal term is the usual phonon group velocity: $v_{\bm q,\nu\nu}=\nabla_{\bm q} \omega_{\bm q \nu}$. These relations allow us to to cast the energy current, Eq. \eqref{eq:J-QHA} into the Hardy form \cite{Hardy_flux,Isaeva2019,Caldarelli2022}:
\begin{multline}
 \widehat J = \hbar\sum_{\bm q \nu \nu'}\frac{\omega_{\bm q \nu}+\omega_{\bm q \nu'}}{2}v_{\bm q\nu\nu'}\ha^\dag_{\bm q\nu} \ha_{\bm q\nu'} \\ ~~ + \hbar\sum_{\bm q \nu \nu'}\frac{\omega_{\bm q \nu}-\omega_{\bm q \nu'}}{4}v_{\bm q \nu \nu'} \bigl ( \ha_{-\bm q\nu}\ha_{\bm q\nu'}-\ha^\dag_{\bm q\nu}\ha^\dag_{-\bm q\nu'} \bigr ).
\end{multline}
% \begin{align}
% J^{\alpha}&=J^{R,\alpha}+J^{A,\alpha}\\
% &=\hbar\sum_{\bm q \nu \nu'}\frac{\omega_{\bm q \nu}+\omega_{\bm q \nu'}}{2}v_{\bm q\nu\nu'}^\alpha\ha^\dag_{\bm q\nu}\ha_{\bm q\nu'}\\
% &+\hbar\sum_{\bm q \nu \nu'}\frac{\omega_{\bm q \nu}-\omega_{\bm q \nu'}}{4}v_{\bm q\nu\nu'}^\alpha[\ha_{-\bm q\nu}\ha_{\bm q\nu'}-\ha^\dag_{+\bm q\nu}\ha^\dag_{-\bm q\nu'}]
% \end{align}
%
% A great, exact, simplification in the crystal case is the block nature of the velocity matrix (no $\bm q \bm q'$ terms). In this basis:
% \( v_{\bm q\nu\nu'}=v_{\bm q\nu'\nu}^*=-v_{-\bm q\nu'\nu}\)
% and the diagonal part $v_{\bm q}^{\nu\nu}=\nabla_{\bm q} \omega_{\bm q \nu}$. 
%
Neglecting again the anti-resonant (second-line) part of the flux, we find:
\begin{multline}\label{eq:GK_manybody_crystal}
 \kappa =\frac{i}{VT}\sum_{\bm q \bm k \nu \nu' \mu \mu'}c_{\bm k \mu \mu'}v_{\bm q\nu\nu'} v_{\bm k\mu\mu'}
\times\\ 
\frac{\omega_{\bm q\nu\nu'}}{\omega_{\bm k\mu\mu'}}\widebar \Lambda^{-1}_{\bm q\nu\nu',\bm k \mu \mu'}(0),
\end{multline}
\resub{where $c_{\bm k\mu\mu'}=c_{\bm k\mu\bm k\mu'}$.}

% \subsection{Boltzmann Transport Equation}
% Let us draw a comparison with BTE. In Eq.~\ref{eq:GK_manybody_crystal} we applied the effects of periodicity, which transform the Dynamical Matrix and consequently the velocity matrix in a block matrix, with a block for each wavevector. That said, to compare our theory to BTE another simplification is needed. BTE is constructed in the quasi-particle framework and consider the velocity of the phonon $\bm v_{\bm q}^\nu$, while both the WTE and QHGK have a matrix definition for the velocity $v_{\bm q\nu\nu'}$, whose diagonal (in the Bloch basis), is the velocity of the phonon. 

\resub{In Appendix \ref{sec:distinct-bands} it is shown}
%Sec.~S-III of the Appendix shows 
that when the phonon line-widths are small with respect inter-band separations, the $\widebar\Lambda$ matrix in Eq. \eqref{eq:GK_manybody_crystal} is diagonal in the $\nu\nu'$ and $\mu\mu'$ indices. Assuming that this is the case, the matrix $\Omega_{IJ}$ in Eqs. \eqref{eq:OmegaIJ} and \eqref{eq:OmegaDiag} vanishes, and Eq. \eqref{eq:GK_manybody_crystal} becomes:
\begin{align}\label{eq:kappa_gk_mb_crystal_no_band}
 \kappa =\frac{1}{V}\sum_{\bm q\nu \bm k \mu}c_{\bm k\mu}v_{\bm q\nu} v_{\bm k \mu} \frac{\omega_{\bm q\nu}}{\omega_{\bm k\mu}}\widebar{\Gamma}^{-1}_{\bm q \nu \bm k \mu}(0),
\end{align}
\resub{where $c_{\bm k\mu}=c_{\bm k\mu\mu}=\frac{n_{\bm k\mu}(n_{\bm k\mu}+1)\hbar^2\omega^2}{k_BT^2}$ is the modal heat capacity.}
If we now define the \resub{scattering} matrix $S$ as
\begin{equation}\label{eq:scattering-matrix}
 S_{\bm q\nu \bm k \mu}=\widebar{\Gamma}_{\bm k \mu \bm q\nu }\frac {\omega_{\bm k \mu}}{\omega_{\bm q \nu}},
\end{equation}
then Eq. \eqref{eq:kappa_gk_mb_crystal_no_band} can be written as:
\begin{align} \label{eq:kappaBTE}
 \kappa =\frac{1}{V}\sum_{\bm q \nu \bm k \mu}c_{\bm q \nu} v_{\bm q \nu} v_{\bm k \mu}S^{-1}_{\bm q\nu\bm k \mu},
 % \nonumber\\
 % &=\sum_{\bm q \nu\bm q' \mu}c_{\bm q \nu} v_{\bm q \nu} v_{\bm q' \mu}\widebar{\Gamma}^{-1}_{\bm q \nu \bm q' \mu}\sqrt{\frac{c_{\bm q' \mu}}{c_{\bm q \nu}}}\nonumber\\
 % &=\sum_{\bm q \nu\bm q' \mu}\sqrt{c_{\bm q \nu} c_{\bm q' \mu}} v_{\bm q \nu} v_{\bm q' \mu}\widebar{\Gamma}^{-1}_{\mu\mu'} \nonumber\\
 % &=\kappa^{QHGK-full}\nonumber
\end{align}
which has the same form as from the full BTE, provided $S$ can be identified with the scattering matrix appearing therein \cite{BTE:Ziman2001,articolo_kaldo}.
% Interestingly, if we apply the same reasoning to QHGK we find that its Markovian limit in a crystal neglecting band interactions is BTE-RTA. Formally, the relation between the QHGK theory and the BTE is the same for both the single and many body theories.
%
% \subsection{Practical example: third order anharmonic potential}
In order to see that this is indeed the case, we compute the memory matrix, Eq. \eqref{eq:Gamma}, to lowest (second) order in the cubic anharmonic interactions, $K$, %Eq. \eqref{eq:hamiltonian_quantum}, 
and taking into account lattice periodicity. 
%\AF{La definizione della parte anarmonica e' solo nell'appendice ora. Dici che va bene?}
%
The calculations are quite lengthy and are fully reported in \resub{Appendix \ref{sec:memory-matrix}}.
%Sec.~S-IV of the Appendix. 
%reported in the \textcolor{red}{supplementary} in details. Here we report the most significant results. We indicate real and imaginary part in the following way: \(\Gamma=\Gamma'+i\Gamma''\)
%
% \subsection{Memory matrix in the Boltzmann case}
%In the Boltzmann framework, due to periodicity and the quasi-particle picture, we impose that both $A_I$ and $A_J$ are composed only by operators with the same wavevector and band index, i.e. same phonon. In other words $(I,J)\to (pp,qq)\to(\bm q \nu\bm q \nu,\bm q' \nu'\bm q' \nu' )$. Thus:
The final, purely real, results reads:
\vskip-35pt
\begin{widetext}
\begin{multline}\label{eq:memory_matrix_bte}
 \widebar{\Gamma}_{\bm q \nu \bm k \mu}(z=0)
 %&=\widebar\Gamma_{\bm q \nu}^0\delta_{\bm q \nuq}+\widebar\Gamma_{\bm q \nuq}^1\\ &
 =2\gamma_{\bm q \nu}\delta_{\bm q \bm k}\delta_{\nu\mu}+\pi\hbar\sum_{\bm q' \nu'}|K_{\bm q \nu \bm k \mu \bm q' \nu' }|^2n_{\bm q' \nu' } \left [ \frac{(n_{\bm q \nu}+1)}{ n_{ \bm k \mu } }\delta(\omega_{\bm q \nu}+\omega_{ \bm k \mu } -\omega_{\bm q' \nu' }) \right . \\ \left .
 -\frac{n_{\bm q \nu} }{n_{ \bm k \mu }} \delta(\omega_{\bm q \nu} +\omega_{\bm q' \nu' }-\omega_{ \bm k \mu } ) -\frac{n_{\bm q \nu}+1 }{n_{ \bm k \mu }+1} \delta(\omega_{ \bm k \mu } +\omega_{\bm q' \nu' }-\omega_{\bm q \nu} ) \right ],
\end{multline}\nopagebreak[4]
where $K_{\bm q \nu \bm k \mu \bm q' \nu' }$ is non-zero only for momentum-conserving triplets and $\gamma_{\bm q \nu}$ is the anharmonic phonon linewidth \cite{FGR_allen}:
\begin{equation}\label{eq:phonon_linewidth}
 \gamma_{\bm q \nu} =\pi\hbar\sum_{\bm q' \nu' \bm k \mu}|K_{\bm q \nu \bm k \mu\bm q' \nu'}|^2 \left [\frac{1}{2}(n_{\bm k \mu}+n_{ \bm q' \nu' }+1)\delta(\omega_{\bm q \nu}-\omega_{\bm k \mu}-\omega_{ \bm q' \nu' }) + (n_{\bm k \mu}-n_{ \bm q' \nu' }) \delta(\omega_{\bm q \nu}+\omega_{ \bm q' \nu' }-\omega_{\bm k \mu})\right ].
\end{equation}
Eqs. \eqref{eq:memory_matrix_bte} and \eqref{eq:phonon_linewidth} coincide with those
%which through Eq.~\ref{eq:scattering-matrix} is related with the expression of the scattering matrix 
appearing in the full BTE, computed to lowest order in the cubic anharmonic corrections to the lattice Hamiltonian \cite{articolo_kaldo}, thus proving the equivalence of the treatments of thermal transport based on the Boltzmann's kinetic approach and the Green-Kubo theory of linear response. 

\begin{resubtext} \section{Comparison with the Wigner Transport Equation} \label{sec:WTE}
As an alternative to the GK theory, the BTE for heat transport can be derived from many-body perturbation theory, leveraging a Wigner-like lattice distribution obtained from phonon Green's functions \cite{Wigner_Horie}. Recently, this approach has been considerably refined by introducing a dependence of the Wigner distribution on band indices, which give rise to inter-band contributions to the heat conductivity when the distance between neighbouring bands is comparable with the phonon linewidth \cite{Simoncelli2019,Caldarelli2022,Simoncelli2021}. The final expression for the heat conductivity in the WTE approach is given by Eq. (12) of Ref.~\cite{Simoncelli2019}, reading: 
%\SBnote{qua la ricopiamo pari-pari con la loro notazione (lasciamo la $V$ maiuscola, perché è diversa dalla nostra, ma mettiamo minuscolo, ad esempio $N$ perché è lo stesso nostro numero di occupazione.}
\begin{multline} \label{eq:WTE}
 \kappa^{WTE} =
 %\frac{1}{V}\Big[\sum_{\bm q \nu \bm k \mu}c_{\bm q \nu} v_{\bm q \nu} v_{\bm k \mu}S^{-1}_{\bm q\nu\bm k \mu} + 
\kappa^{BTE} + \\ \frac{\hbar^2}{k_bT^2V}
\sum_{\bm{q} \nu\neq \nu'} \frac{\omega_{\bm q \nu'}+\omega_{\bm q \nu}}{4}(\omega_{\bm q \nu}n_{\bm q \nu}(n_{\bm q \nu}+1)+\omega_{\bm q \nu'}n_{\bm q \nu'}(n_{\bm q \nu'}+1)) V_{\bm q\nu\nu'} V_{\bm q\nu'\nu} \frac{\gamma_{\bm q \nu}+\gamma_{\bm q \nu'}}{(\omega_{\bm q \nu}-\omega_{\bm q \nu'})^2+(\gamma_{\bm q \nu}+\gamma_{\bm q \nu'})^2},
 \end{multline}
 \resub{where $\kappa^{BTE}$ is the BTE expression for the heat conductivity given by Eq. \eqref{eq:kappaBTE} and $V_{\bm q\nu\nu'} = \frac{2\sqrt{\omega_{\bm q \nu}\omega_{\bm q \nu'}}}{\omega_{\bm q \nu}+\omega_{\bm q \nu'}}v_{\bm q\nu'\nu}$ is the velocity matrix defined in that work. By treating the difference between Eq. \eqref{eq:GK_manybody_crystal} and Eq. \eqref{eq:kappaBTE} in the RTA, the former can be cast into the form:}
 %written with our and Hardy\cite{Hardy_flux} notation of generalized group velocity to allow comparison to our formulas. The two velocities are related by $2\sqrt{\omega_{\bm q \nu}\omega_{\bm q \nu'}}v_{\bm q\nu'\nu}=(\omega_{\bm q \nu}+\omega_{\bm q \nu'})v^{WTE}_{\bm q\nu'\nu}$\cite{Caldarelli2022}.
%
%We already proved in Eq.~\ref{eq:QHGK_limit}-\ref{eq:k_markovian} that neglecting vertex corrections we can treat such materials as QHGK \cite{Isaeva2019} does. However, we can retain a certain degree of vertex corrections following the strategy in Ref.~ \cite{Simoncelli2019,simoncelli-thesis}. Let us treat in the single mode RTA approximation the interband terms only. This leads to:
\begin{equation}\label{eq:I-QHGK}
 \kappa =
 %\frac{1}{V}\Big[\sum_{\bm q \nu \bm k \mu}c_{\bm q \nu} v_{\bm q \nu} v_{\bm k \mu}S^{-1}_{\bm q\nu\bm k \mu} 
 \kappa^{BTE} +
 \frac{1}{V}\sum_{\bm q \nu\neq \nu'}c_{\bm q \nu \nu'}v_{\bm q\nu\nu'}v_{\bm q\nu'\nu}\frac{\gamma_{\bm q \nu}+\gamma_{\bm q \nu'}}{(\omega_{\bm q \nu}-\omega_{\bm q \nu'})^2+(\gamma_{\bm q \nu}+\gamma_{\bm q\nu'})^2},
\end{equation}
\end{resubtext}
\end{widetext}
% \SBnote{E' questo che vogliamo dire? Se SI, spiega meglio. Se NO, correggi ...}\AFnote{Mi sembra proprio quello che volevamo dire. Vuoi che lo commenti di piu'?} 
which differs from Eq.~\eqref{eq:WTE} only by corrections of order $\mathcal O(\gamma^2/\omega^2)$. 
% This approximation of full QHGK 
%, here labelled Intermediate-QHGK (I-QHGK), 
% greatly reduces the computational cost in respect of Eq.~\ref{eq:GK_manybody_crystal} since does not require the computation of a larger scattering matrix. 
These considerations show that,
% Moreover, its relation to WTE shows that 
provided the same levels of approximation \resub{are adopted}, the WTE and GKMZ give the same results, in line with the work of Ref.~ \cite{Caldarelli2022} and previous work for the electrical conductivity as reported, \emph{e.g.}, in Mahan's textbook \cite{Mahan1990}. \begin{resubtext}Crucially, a proper account of inter-band contributions to heat conduction allows the WTE to be easily generalized to disordered systems, which, strictly speaking, do not display any dispersions, as they lack translational symmetry. This result emerges naturally from the GK approach, which does not presuppose any symmetry.

We conclude that crystals whose interband spacing is comparable with their phonons’ linewidths can not be described by full BTE alone. Among these crystals, we find materials such as (anti-) perovskites with promising applications in various fields of electronics, which will be the subject of a numerical application in Sec. \ref{sec:Li3ClO}. 

\begin{figure}[h!]
 \centering
 \includegraphics[width=0.5\textwidth]{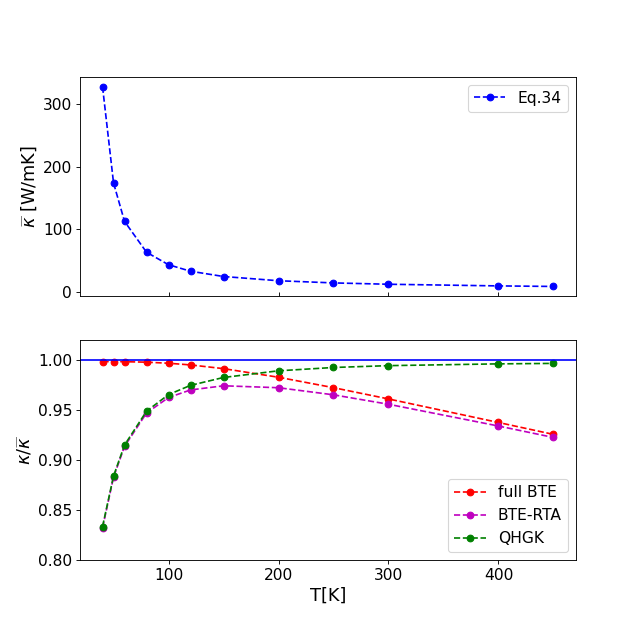}
 \caption{Top: thermal conductivity averaged over the three Cartesian coordinates as a function of temperature, computed with Eq.~\eqref{eq:I-QHGK}. Bottom: ratio between the thermal conductivity computed with different methods and the one in the top panel. }
 \label{fig:comparison_method}
\end{figure}

\section{Application to $\mathbf{Li_3ClO}$} \label{sec:Li3ClO}
The theory presented in this paper has been demonstrated on the lithium-rich anti-perovskite Li$_3$ClO, which is a promising candidate for all-solid-state lithium-metal batteries and whose  transport properties have been recently studied with state-of-the-art methods \cite{Pegolo2022}. We computed the heat conductivity of this material using the present theory and the technical details reported in Appendix \ref{sec:material_simulation}. In Fig.~\ref{fig:comparison_method} we display our results obtained using different approximations. We identify two distinct regimes. At low temperature, hydrodynamic effects \cite{Cepellotti2015}---which are accounted for in the full BTE but not in QHGK---may considerably enhance the heat conductivity, while interband contributions in Eqs. (\ref{eq:WTE}-\ref{eq:I-QHGK})---present in QHGK but not in the BTE---are negligible due to the vanishing of the vibrational linewidths as $T\to 0$; in this regime, the full BTE correctly describes the temperature dependence of the heat conductivity, while QHGK does not. At room temperature and above, instead, interband effects may be important, particularly in complex crystals as in the present case, while hydrodynamic effects are negligible; in this case, the QHGK approximation is applicable and the BTE is not. Eqs. (\ref{eq:WTE}-\ref{eq:I-QHGK}) nicely and correctly interpolate between these two regimes. 

% . We conclude that at low temperature the full BTE---which account for hydrodymamic effects but not for inter-band contributions---

% which is in fact underestimated in the QHGK and BTE-RTA. At room temperature and above, the single-mode RTA approximation is valid: the BTE and QHGK methods differ only for the interband contributions i.e. the second part of Eq.~\eqref{eq:I-QHGK}. At low temperatures, the single-mode approximation cannot describe the hydrodynamic effects\cite{Cepellotti2015}, while the interband contributions are negligible. Indeed, the interband contributions matter when the distance between bands is in the order of a few linewidths (see Appendix.~\ref{sec:material_simulation} for details), condition no longer satisfied at low temperatures, since the anharmonic linewidths decrease with temperature. Eq.~\ref{eq:I-QHGK} captures both of these regimes since it contains both the interband term and the BTE scattering matrix. Given the separation at low and high temperatures of these effects: the hydrodynamic effect/failure of single-mode approximation and the interband one, Eq.~\eqref{eq:I-QHGK} seems a legit approximation far less expensive than full QHGK. The necessity of full QHGK, Eq.~\eqref{eq:GK_manybody_crystal}, would arise only if both the single-mode RTA approximation and the intraband approximation were to fail. 

\end{resubtext}

% \begin{figure}[H]
% \centering
% \includegraphics[width=0.5\textwidth]{plots/ratio_methods.png}
% \caption{Caption}
% \label{fig:ratio_method}
% \end{figure}

% Let us compare our formula to the one in Ref.~ \cite{articolo_kaldo}. Associating $K_{pqr}$ to their $\phi_{\bm q\bm q'\bm q''}^{\nu \nu'\nu''}\delta_{\bm q+\bm q'+\bm q'',G}$ so to conserve crystal momentum \emph{mod} a generic reciprocal lattice $\bm G$, we indeed obtain that:
% \(
% \Gamma_{\bm q\nu \bm q' \mu}^{S}=\widebar{\Gamma}_{\bm q\nu \bm q' \mu}\sqrt{\frac{c_{\bm q' \mu}}{c_{\bm q \nu}}},
% \)
% proving that for~ this anharmonic potential BTE-full e is just a special case of a more general Many Body theory. 

% \section{Conclusions }\label{sec:Conclusions}
%The main goal of this paper was obtaining a general Green-Kubo formalism to treat thermal conduction in insulators far from the point of melting.
\section{Conclusions} \label{sec:conclusions}
In this paper we have critically analyzed the approximations that lead to the unified quasi-harmonic Green-Kubo approach to heat transport in crystalline and disordered insulators \cite{Isaeva2019} and shown how they can be dealt with by applying the Mori-Zwanzig memory-function formalism to one- and two-normal-mode correlation functions. In the first case, the MZ formalism allows one to account for vibrational memory (\emph{self-energy}, in the many-body parlance) effects on heat conduction, whereas in the second it permits to dispose of the \emph{dressed-bubble} approximation to the two-mode correlation functions and derive an expression for the heat conductivity that is equivalent to that provided by the full linearized Boltzmann's transport equation. Besides providing a fully \emph{ab initio} derivation of the latter, we believe that our paper will pave the way to the study of systems where memory effects and vertex corrections to heat transport coefficients are both important. \resub{Moreover, we have clarified the interconnection and equivalence between the Wigner and quasi-harmonic Green-Kubo approaches to heat transport in solids, provided the same levels of approximation are adopted---thus extending the work of Ref.~ \cite{Caldarelli2022}. }

\begin{acknowledgments}
 We are grateful to Giuseppe Barbalinardo, Enrico Drigo, and Paolo Pegolo for many insightful discussions and suggestions. This work was partially funded by the EU through the \textsc{MaX} Centre of Excellence for supercomputing applications (grant No. 824143) and by the Italian MUR, through the PRIN grant \emph{FERMAT} (grant No. 2017KFY7XF)
\end{acknowledgments}

\appendix
\section{Derivation of Eq. (\ref{eq:QHGK_limit})} \label{sec:S-MemoryDiagonal}
In order to prove Eq. \eqref{eq:QHGK_limit}, we first write Eqs. (\ref{eq:GK_manybody}-\ref{eq:CIJ(0)}) in the diagonal ($I=J$) approximation:
\begin{align} \label{eq:kappa-diag}
\kappa &\approx \frac{\hbar^2}{VT}\sum_{pq}v_{pq}v_{pq} \widebar\omega_{pq}^2 \int_0^\infty \bigl (\widehat A_{pq}(t), \widehat A_{pq} \bigr ).
% % \widebar \Lambda_{pq,pq}(0) &= \omega_q-\omega_p +i \Gamma_{pq,pq}(0), \\
% % \Gamma_{pq,pq}(0) &\doteq \Gamma'_{pq}+i\Gamma''_{pq} \\
% % & = ...
\end{align}
We now observe that $\widehat A_{pq}=\widehat A_{qp}^\dagger$ and that the Kubo inner product, Eq. \eqref{eq:scalar_product}, has the property: $\bigl (\widehat A,\widehat B \bigr ) = \bigl (\widehat A^\dagger, \widehat B^\dagger)^*$. We conclude that:
\begin{equation}
 \int_0^\infty \bigl (\widehat A_{qp}(t), \widehat A_{qp} \bigr ) = \left (\int_0^\infty \bigl (\widehat A_{pq}(t), \widehat A_{pq} \bigr ) \right )^*,
\end{equation}
%and that therefore only the real part of the integrals in Eq. \eqref{eq:kappa-diag} contribute to the conductivity, 
from which we deduce that the $\widebar C_{pq,pq}(0)$ (Eq. \ref{eq:CIJdef})
%and $\widebar \Lambda_{pq,pq}(0)$ (Eq. \ref{eq:Lambda})
matrix is Hermitian with respect to $p\leftrightarrow q$ exchange, whereas $\widebar \Gamma_{pq,pq}$ and $\Omega_{pq,pq}$ (Eqs. \ref{eq:Gamma}-\ref{eq:OmegaDiag}) are anti-Hermitian, and Eq, \eqref{eq:QHGK_limit} follows immediately.

\section{Derivation of Eq. 
(\ref{eq:gk_product_green_f})} \label{sec:diag->QHGK}
%As explained in the main text, this is not a one-to-one relation, since there are vertex corrections. Actually the same reasoning is valid for BTE-Full and BTE-RTA. Looking at Eq.~\ref{eq:memory_matrix_bte}. The diagonal part of the scattering matrix is $2\gamma_{\bm{q}\nu}+\Gamma_{\mathfb q\nu \bm q\nu}^1$, while RTA consider only the lifetimes $1/(2\gamma_{\bm{q}\nu})$. However, in this case such correction $\Gamma_{\mathfb q\nu \bm q\nu}^1$, that we identify as the diagonal vertex correction, is of order $O(1/N_{modes})$ in comparison of $\gamma_{\bm q \nu}$: the former is a sum over $N_{modes}$, the latter a sum over $N_{modes}^2$ terms.
%
%To summarize, in general the diagonal of QHGK-full differs from QHGK, but for this anharmonic potential the difference is negligible in the thermodynamic limit $1/N_{modes}$.
In order to compare Eqs.~\eqref{eq:QHGK_limit} and \eqref{eq:gk_product_green_f}, we first compute the diagonal part $\bigl (II = (pq,pq) \bigr )$ of the memory matrix, \resub{ assuming the anharmonic terms in the vibrational Hamiltonian can be truncated to third order in the atomic displacements:
\begin{equation}\label{eq:hamiltonian_quantum}
% \widehat H= \widehat H^\circ +
 \widehat{\mathcal V} \approx
 % \frac{\hbar^{\frac{3}{2}}}{6}
 \frac{\hbar^{3/2}}{6} \sum_{pqr}K_{pqr}\hx_p\hx_q\hx_r, 
\end{equation}
where 
% $\widehat H^\circ = \sum_p \hbar \omega_p (\widehat a_p^\dagger\widehat a_p+\sfrac{1}{2} )$, 
$\widehat X_p=\had_p+\ha_p$ is a (rescaled) normal-mode coordinate.} Using the method explained in Sec.~\ref{sec:memory-matrix}, we obtain: %and reminding that $C_{pq,pq}=\frac{n_p-n_q}{\hbar(\omega_q-\omega_p)}=n_p(n_q+1)\frac{e^{\beta\hbar(\omega_p-\omega_q)}-1}{\omega_p-\omega_q}$:
\begin{multline}
 \widebar{\Gamma}_{pq,pq}^{'}(z=0)= \\ \widebar{\gamma}_p^{>'}(\omega_q)d_{pq}+\widebar{\gamma}_q^{<'}(\omega_p)d_{qp}+\mathcal O(N^{-1}),
\end{multline}
%
% \begin{multline}
% \widebar{\Gamma}_{pq,pq}^{'}(z=0)=\frac{1}{2}\widebar{\Gamma}_{pp,pp}^{'}(\omega_q-\omega_p)o_{pq}
% +\frac{1}{2}\widebar{\Gamma}_{qq,qq}^{'}(\omega_p-\omega_q)o_{qp} \\ -\frac{\beta\pi\hbar}{2}\sum_{r}|K_{pqr}|^2n_r(n_r+1)\hbar\omega_r\times
% \big[~
% \delta(+\omega_p-\omega_q+\omega_r) +\delta(+\omega_q-\omega_p+\omega_r)\big]
% \end{multline}
% where $o_{pq}=SBfrac{\beta\hbar(\omega_p-\omega_q)}{e^{\beta\hbar(\omega_p-\omega_q)}-1}$.% is the factor due the harmonic $C_{pq}^0(t=0)$ approximation.%, and it differs from $1$ only when $\beta|\omega_p-\omega_q|\gtrsim 1$. 
 where $N$ is the number of normal modes, $(\cdot)'=\mathsf{Re} (\cdot)$, $d_{pq} =\frac{\beta\hbar(\omega_p-\omega_q)}{e^{\beta\hbar(\omega_p-\omega_q)}-1}$, and $\widebar \gamma_p^{\scriptscriptstyle\lessgtr}(\omega)$ are the 
 % memory-function spectra 
FL transforms of the one-body greater/lesser Green's functions, whose expressions as obtained from Mori's formalism \cite{Mori_1965} is
\begin{align}
 \begin{aligned}
 \widebar g_p^>(\omega) &= -i\langle\ha_p\had_p \rangle \frac{1}{(\omega_p-\omega) -i\widebar{\gamma}_p^>(\omega)}, \\
 \widebar g_p^<(\omega) &=i\langle \had_p \ha_p \rangle \frac{1}{(\omega_p-\omega)-i\widebar{\gamma}_p^<(\omega)}.
 \end{aligned}
 \label{eq:gbar>}
\end{align}
% or as FT, taking twice the imaginary part:
The corresponding Fourier transforms equal twice the imaginary part of Eqs. \eqref{eq:gbar>}, reading: 
\begin{align}\label{eq:def_gf_mori}
 \tilde g_p^>(\omega) &=-i\langle\ha_p\ha_p^\dag\rangle\frac{2\widebar \gamma_p^{>'}(\omega)}{|(\omega_p-\omega)-i\widebar{\gamma}_p^>(\omega)|^2}, \\
 \tilde g_p^<(\omega) &=\phantom{-}i\langle\ha_p^\dag\ha_p\rangle \frac{2\widebar \gamma_p^{<'}(\omega)}{|(\omega_p-\omega)-i\widebar{\gamma}_p^<(\omega)|^2} .
\end{align}
 To lowest order in the cubic anharmonic correction to the lattice potential energy, Eq. \eqref{eq:def_gf_mori} reads:
 \begin{widetext}
 \begin{multline}
 \tilde \gamma_p^{<'}(\omega) = \frac{\pi\hbar}{16n_p}\sum_{qr}|K_{pqr}|^2\Big [n_qn_r \delta(\omega-\omega_q-\omega_r)
 +(n_q+1)n_r \delta(\omega+\omega_q-\omega_r) \\ +n_q(n_r+1) \delta(\omega-\omega_q+\omega_r)
 + (n_q+1)(n_r+1) \delta(\omega+\omega_q+\omega_r)\Big ] +\mathcal O(K^3).
 \end{multline}
%
%Let us simplify the expression starting from $\widebar{\Gamma}_{pp,pp}$ definition in Eq.~\ref{eq:memory_matrix_bte}.
Neglecting corrections of order $\mathcal O(N^{-1})$, we obtain:
%consisting in sum over $N_{modes}$ in presence of sum of $N_{modes}^2$ similar terms, we obtain after some manipulation that at the dominant order
\begin{align}
 \kappa&=\frac{1}{ V}\sum_{pq}\frac{\hbar^2(\omega_p+\omega_q)^2}{4}v_{pq} v_{pq} (I_{pq}+I_{qp}), \text{ where} \\
 I_{pq}&=\frac{n_p(n_q+1)}{2k_bT^2} \frac{\widebar\gamma_p^{>'}(\omega_q)+\widebar\gamma_q^{<'}(\omega_p)}{|(\omega_p-\omega_q)-i(\widebar\gamma_p^>(\omega_q)d_{pq}+\widebar\gamma_q^<(\omega_p)d_{qp})|^2}. \label{eq:I_pq_from_manybody}
\end{align}
%
%where the greater and lesser green's function have also been solved inside Mori Theory with the scalar product $\langle A|B\rangle=\langle A^\dag B\rangle$ in its scalar form i.e. $N_{modes}$ independent Memory function, one for each mode $p$.

Let us compare this expression with Eq.~\eqref{eq:gk_product_green_f}, which gives the QHGK approximation within the full dressed-bubble approximation, including memory (non-Markovian) effects. The integral in Eq.~\eqref{eq:gk_product_green_f} can be computed using Cauchy's residue theorem. 
%Let us assume $\omega_q\neq \omega_p$, leaving the degenerate case $\omega_p=\omega_q$ to the reader. 
For instance, in the $\omega_q\neq \omega_p$ case, substituting Eq.~\eqref{eq:def_gf_mori} into Eq.~\eqref{eq:integral_overlap_spectrum}, we obtain an integrand with $4$ distinct poles, whose positions in the QH approximation are:
\begin{align}
 \begin{aligned}
 \omega_p^{\pm}=\omega_p+\widebar{\gamma}_p^{''}(\omega_p)\pm i\widebar{\gamma}_p^{'}(\omega_p)+ \mathcal O(\gamma_p^2)\\
 \omega_q^{\pm}=\omega_q+\widebar{\gamma}_q^{''}(\omega_q)\pm i\widebar{\gamma}_q^{'}(\omega_q)+ \mathcal O(\gamma_q^2),
 \end{aligned}
\end{align}
where $\widebar{\gamma}_q^{''} = \mathsf{Im}\widebar\gamma_q$.
Closing the path in the upper complex half-plane, we get:
\begin{align}\label{eq:I_pq_with_residue}
 I_{pq} &= 2\pi i\sum_{\mathrm{Im}\widebar{z}>0} \mathsf{Res}(f(\widebar{z})) \\
 &= \frac{(n_p+1)n_q}{2k_bT^2}
 \left(\frac{\widebar \gamma_p^{>'}(\omega_q^+)}{|(\omega_p-\omega_q^+)-i\widebar \gamma_p^>(\omega_q^+)|^2} +\frac{\widebar \gamma_q^{<'}(\omega_p^+)}{|(\omega_q-\omega_p^+)-i\widebar \gamma_q^>(\omega_p^+)|^2} \right)\nonumber,
\end{align}
\end{widetext}
where we used $\langle \ha_p\ha_p^\dag\rangle\approx n_p+1$. We observe that Eq.~\eqref{eq:I_pq_with_residue} and Eq.~\eqref{eq:I_pq_from_manybody} have the same numerator but a slightly different denominator. Their difference is negligible if $\left| \frac{\widebar\gamma_p(\omega_q) -\widebar\gamma_p(\omega_p)}{\omega_p-\omega_q}\right|\ll 1$, which happens in the quasi-harmonic limit $\Bigl (\widebar \gamma,\frac{\partial \widebar\gamma}{\partial \omega}\sim \mathcal O(K^2) \to 0 \Bigr )$ if the memory function is regular enough. If $\widebar\gamma(\omega)=\gamma$, independent of $\omega$, both Eqs.~\eqref{eq:I_pq_with_residue} and \eqref{eq:I_pq_from_manybody} return the Markovian approximation, Eq.~\eqref{eq:k_markovian}, to order $\mathcal O(\gamma^2)$.

%We also notice a difference between $c_{pq}$ and $\frac{(\omega_p+\omega_q)^2}{4} \frac{n_p(n_q+1)+n_q(n_p+1)}{2k_bT^2}$. Such difference goes to zero if $k_bT\gg\hbar |\widebar \gamma_p(\omega_q)+\widebar \gamma_q(\omega_p)|$, since only couples such that $|\omega_p-\omega_q|\sim|\widebar \gamma_p'(\omega_q)\widebar \gamma_q(\omega_p)|$ give a significant $I_{pq}$. In such approximation $o_{pq}\approx 1$.
%
% To summarize, at the dominant order
% FUll-QHGK reproduces the Beyond Markovian QHGK, at least for this perturbation.
In these calculations we have neglected terms of order $\mathcal O(N^{-1})$, which can be identified with the diagonal part of the vertex corrections. We notice that similar $\mathcal O(N^{-1})$ corrections are also neglected when comparing BTE-RTA, $\tau_{\bm q \nu}=1/2\gamma_{\bm q \nu}$, with the diagonal of the full BTE scattering matrix: $S_{\bm q \nu \bm q \nu}= 2\gamma_{\bm q \nu}+ \mathcal O(N^{-1})$.

\section{Derivation Eq.~(\ref{eq:kappa_gk_mb_crystal_no_band}) for well separated bands} \label{sec:distinct-bands}
Eq. \eqref{eq:kappa_gk_mb_crystal_no_band} is derived under the assumption the the inter-band ($\nu\ne\nu'$ and $\mu\ne\mu'$) elements of the $\widebar\Gamma^{-1}_{\bm q\nu\nu',\bm q'\mu\mu'}$ matrix can be neglected. This is indeed the case when the phonon bands are well separated, in the sense that $|\omega_{\bm q\nu}-\omega_{\bm q\nu'}|\gg \Gamma_{\bm q\nu\nu',\bm k\mu\mu'} ~\forall~(\bm k\mu\mu')$. We say in this case that \emph{interband contributions} are negligible and we call the corresponding approximation the \emph{intraband approximation}. 
%We use the matrix identity:
% \begin{equation}
% (X+Y)^{-1}=X^{-1}-X^{-1}Y(X+Y)^{-1} .
% \end{equation}
%We now 
Let us decompose the $\widebar \Lambda$ matrix into a diagonal part, $D_{IJ}=(-\Omega_{II}+i\widebar \Gamma_{II})\delta_{IJ}$, and an off-diagonal part $O=\widebar\Lambda-D$ and apply the identity:
\begin{equation}
 \widebar \Lambda^{-1}=(D+O)^{-1}=D^{-1}-D^{-1}O(D+O)^{-1}.
\end{equation}
The last term can be rewritten as:
\begin{widetext}
\begin{equation}
 \bigl (D^{-1}O(D+O)^{-1}) \bigr )_{IJ}=\sum_K D^{-1}_{IK}\bigl (O(D+O)^{-1} \bigr )_{KJ}=D^{-1}_{II}\bigl (O(D+O)^{-1} \bigr )_{IJ}.
\end{equation} 
The $O$ matrix does not diverge when $ {|\omega_{\bm q \nu}-\omega_{\bm q \nu'} |\to \infty}$. Therefore,
\begin{equation}
 \lim_{\substack{|\omega_{\bm q \nu}-\omega_{\bm q \nu'} |\to \infty 
 % \\ \nu\neq \nu'
 }} \left [ \frac{1}{\omega_{\bm q\nu'}-\omega_{\bm q\nu}+i\widebar \Gamma_{\bm q \nu \nu',\bm q \nu \nu'}}-\frac{1}{\omega_{\bm q\nu'}-\omega_{\bm q\nu}+i\widebar \Gamma_{\bm q \nu \nu',\bm q \nu \nu'}}\bigl (O(D+O)^{-1} \bigr )_{\bm q \nu \nu',\bm k \mu \mu'} \right ] = 0.
\end{equation}
\end{widetext}
Therefore, only the $\nu=\nu'$ elements survive in this limit. Due to the relation between $\widebar \Lambda_{IJ}$ and $\widebar \Lambda_{JI}$, the argument can be repeated for the second pair of band indices $\mu,\mu'$. Thus, for well separated bands:
\begin{equation}
 (\widebar \Lambda)^{-1}_{\bm q \nu \nu',\bm k \mu \mu'}\approx \delta_{\nu\nu'}\delta_{\mu\mu'}(\widebar \Lambda)^{-1}_{\bm q \nu \nu,\bm k \mu \mu},
\end{equation}
which motivates the \emph{intraband approximation}.
%For the diagonal case $I=J$ is evident that when we consider bands far away, $D^{-1}_{\bm q \nu\neq \nu',\bm q \nu \neq\nu'}\to 0$. Since we know for hypothesis that $O$ (which contains only $\Gamma$ terms) is much smaller the inter-band distances we expect the diagonal interband contributes to decrease with the distance of bands, as in the QHGK case. Let us consider the off-diagonal $I\neq J$case:

% \(
% (D+O)^{-1}_{\bm q \nu \nu',\bm k \mu\mu'}=0-\frac{1}{\omega_{\bm q\nu'}-\omega_{\bm q\nu}+i\widebar \Gamma_{\bm q \nu \nu',\bm q \nu \nu'}}(O((D+O)^{-1}))_{\bm q \nu \nu',\bm k \mu \mu'}
% \)
% The denominator in the expression is far greater for $\nu\neq\nu'$ than in the $\nu=\nu'$ case for hypothesis. Therefore, it is reasonable to assume that the $\nu\neq\nu' $ elements of $\Lambda$ can be neglected. Given the connection between $\Lambda_{IJ}$ and $\Lambda_{JI}$, the argumentations is valid also for the second couple of band indeces $\mu\mu'$.

% To summarize, both QHGK and full-QHGK differ respectively by BTE-RTA and full BTE for \emph{interband contributes}, but in both cases such contributes are expected to decay when the bands are far enough from each other.

\section{Computation of the memory matrix in the cubic approximation (Eq. \ref{eq:memory_matrix_bte})} \label{sec:memory-matrix}

In order to compute the memory matrix leading to the BTE, Eq. \eqref{eq:memory_matrix_bte}, we express the $\widebar \Gamma_{IJ}$ matrix in Eq. \eqref{eq:Gamma} as the FL transform of the time-correlation function of the projected time-derivatives of the $\widehat A_I$ and $\widehat A_J$ operators:
\begin{equation}
 \widebar \Gamma_{IJ}(0) =\frac{1}{C_{JJ}^\circ} \int_0^\infty \Bigl (\Q\dot{\widehat A}_I, \mathrm{e}^{-i\Q\L\Q t}\Q\dot{\widehat A}_J \Bigr ) dt,
\end{equation}
where we used the harmonic approximation of $C_{IJ}$ and
%In this section we show how to compute the Memory Matrix. Eq.~\ref{eq:Gamma} tells that to obtain $\widebar \Gamma$ we need to compute the Laplace-Transform of $\widebar \Gamma$ at $z=0$. $\widebar \Gamma(t)$ is proportional to the correlation of the orthogonal time derivative of $A_I,A_j$:
\begin{widetext}
\begin{align}\label{eq:orthogonal_time_derivative}
 \Q \dot{\widehat A}_I\doteq \frac{i}{\hbar} \Q [ \had{a}_p\ha_q,\widehat H] =-\frac{i\hbar^{1/2}}{2}\sum_{rs}K_{prs}\hx_r\hx_s\ha_q+\ha^\dag_p\frac{i\hbar^{1/2}}{2}\sum_{rs}K_{qrs}\hx_r\hx_s,
\end{align}
\end{widetext}
\resub{where we use a cubic anharmonic potential as in Eq. ~\ref{eq:hamiltonian_quantum}.}
As explained in the main text, at our desidered order of approximation, $\widebar \Gamma\sim \mathcal O( K^2)$, both the average and the Liouvillian operator can be evaluated in the harmonic approximation.

Regarding the FL transforms (LFT) ($\widebar f(\omega)$) it is computed through the Fourier Transform (FT) ($\tilde f(\omega)$ ), using the relation:
\begin{align}\label{eq:LFT_identity}
 \widebar f(\omega)&=\int_0^\infty e^{i\omega t} f(t)
 %&=\int_{-\infty}^\infty e^{i\omega t}\theta(t) f(t)\\
 % &=\frac{1}{2\pi}\int_{-\infty}^\infty d\omega~ \theta(\omega-\omega')f(\omega')\nonumber\\&
 \\ &=\frac{1}{2} \Tilde f(\omega)+\frac{i}{2\pi}\int_{-\infty}^\infty d\omega~\frac{1}{\omega-\omega'} \tilde f(\omega').
% \nonumber
\end{align}
% where $\theta(t)$ is the Heaviside function. 

Assuming that the FT is real, half of it is the real part of the LFT (the dissipative part), while
%. On the other hand, 
the imaginary part would be given by the second term in the relation, the convolution one.

Combining Eq.~\eqref{eq:Gamma} and Eq.~\eqref{eq:orthogonal_time_derivative} we obtain:
\begin{widetext}
\begin{equation}\label{eq:gamma_t_anharmonic}
\begin{aligned}
 \widebar \Gamma'_{pq,rs}(0) &= \frac{\hbar}{8C_{rs,rs}}\sum_{tuvz} \int_{-\infty}^\infty dt\int_0^\frac{1}{k_bT}d\lambda \times~ \\
 &\quad \left[
 K_{ptu} K_{rvz}\left\langle \hx_t(\tau) \hx_u(\tau) \ha_q(\tau) \ha_s^\dag\hx_v\hx_z\right\rangle^\circ + K_{qtu} K_{rvz} \left\langle \ha_p^\dag(\tau)\hx_t(\tau)\hx_u(\tau)\hx_v\hx_z\ha_r\right\rangle^\circ \right . \\ &\left . \quad -K_{ptu} K_{rvz}\left\langle \hx_t(\tau)\hx_u(\tau)\ha_q(\tau)\hx_v\hx_z \ha_r\right\rangle^\circ -K_{qtu} K_{rvz}\left\langle \ha_p^\dag(\tau)\hx_t(\tau)\hx_u(\tau)\ha_s^\dag\hx_v\hx_z\right\rangle^\circ\right ].
\end{aligned}
\end{equation}
This expression contains a great number of combinations of annihilation/creation operator. Let us compute one of them. For instance:
\begin{align}
 \int_{-\infty}^\infty dt~ \int_0^\beta d\lambda~
 K_{ptu} K_{rvz} \left\langle \ha_t^\dag(\tau)\ha_u^\dag(\tau) \right . & \left . \ha_q(\tau)\ha_s^\dag\ha_v\ha_z\right\rangle^\circ \nonumber \\ & = 2\pi K_{ptu} K_{rvz}\frac{e^{\beta \hbar(\omega_t+\omega_u-\omega_q)}-1}{\hbar(\omega_t+\omega_u-\omega_q)}\delta(\omega_t+\omega_u-\omega_q) \langle \ha_t^\dag\ha_u^\dag\ha_q\ha_s^\dag\ha_v\ha_z\rangle^\circ \nonumber \\
 &=2\pi \beta K_{ptu} K_{rvz}\delta(\omega_t+\omega_u-\omega_q)\langle \ha_t^\dag\ha_u^\dag\ha_q\ha_s^\dag\ha_v\ha_z\rangle^\circ.
\end{align}
\end{widetext}
% where we factorized the time evolution and used the property of the $2\pi\delta(\omega)=FT[e^{i\omega t}]$ in the last passage. 
Now we should apply Wick's Theorem on the thermal average, however several of the resulting terms vanish because they are proportional to a delta function of a finite argument. For instance,
%some combinations are avoidable: they would lead to delta functions impossible to satisfy. Ex. 
coupling $\ha^\dag_u\ha_q$ would lead to $\sim \delta(\omega_t)$. Thus, keeping only the non-vanishing combinations, one gets:
\begin{multline}
 2\pi\beta K_{ptu}K_{rvz} (n_tn_u(n_q+1) \\ \times \delta(\omega_t+\omega_u-\omega_q) (\delta_{tv}\delta_{uz} \delta_{qs}+\delta_{tz} \delta_{uv}\delta_{qs})).
\end{multline}
All the other elements of $\widebar\Gamma$ can be computed by performing analogous calculations. We note that, by applying the same argument as in Sec. \ref{sec:S-MemoryDiagonal}, we can conclude that $\widebar\Gamma_{pq,rs}(0)$ is real when $\widehat A_{pq}$ and $\widehat A_{rs}$ are Hermitian.

\begin{figure}[H]
 \centering
 \includegraphics[width=0.5\textwidth]{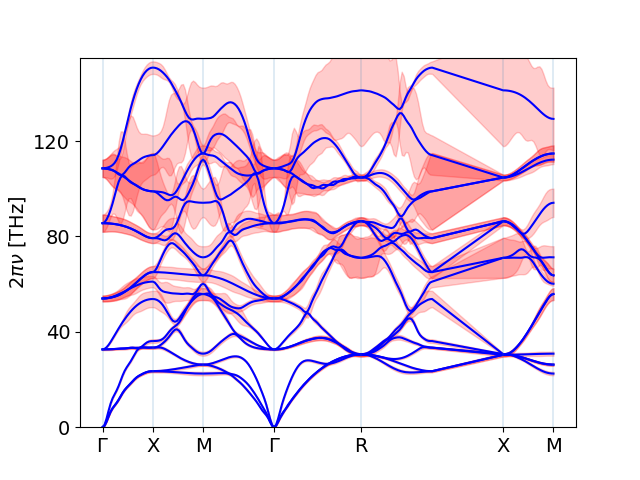}
 \vskip-10pt
 \includegraphics[width=0.45\textwidth]{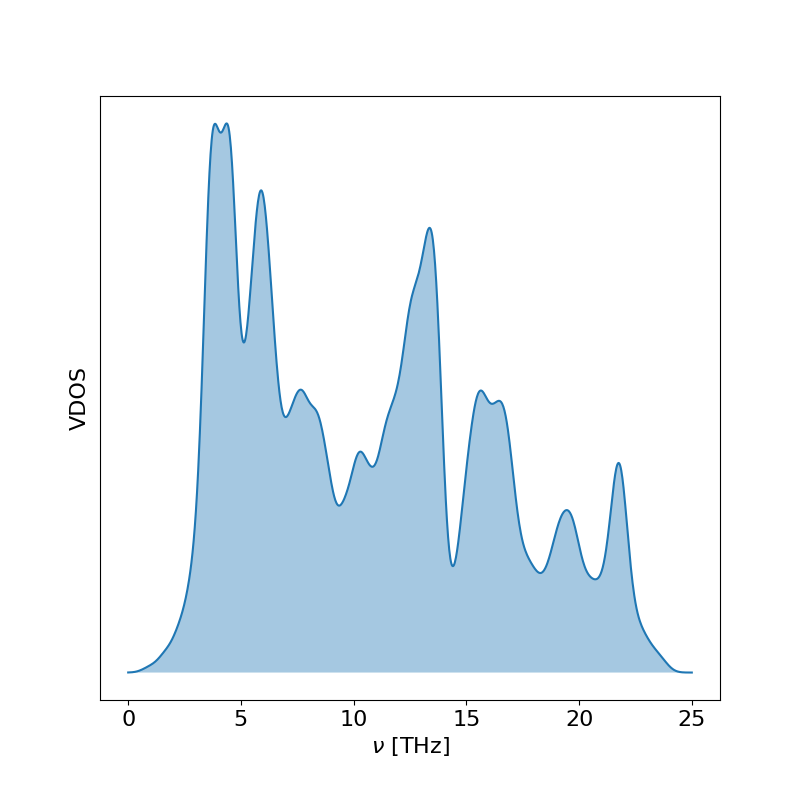}
 \caption{Phonon dispersions and linewidths (upper panel) and VDOS (lower panel) of Li$_3$ClO (see text). }
 \label{fig:VDOS}
\end{figure}

\resub{
\section{Technical details and intermediate results of the numerical application} \label{sec:material_simulation}
The Li$_3$ClO compound is simulated in a cubic cell with edge $a_0=3.875$ \angstrom, using the Buckingham potential \cite{Buck_potential}, and the PPPM \cite{PPPM_Hockney_1988} method to treat the Coulomb interaction. Second- and third-order interatomic force constants where computed with LAMMPS \cite{Lammps} using a finite-difference method in a $[5,5,5]$ supercell, while the $\kappa$aldo code \cite{articolo_kaldo} was used to evaluate vibrational frequencies, anharmonic linewidths, and thermal conductivities. 
% Among the two methods to compute the full BTE in $\kappa$aldo we use the self-consistent one\cite{articolo_kaldo}. Eq.~\eqref{eq:I-QHGK} can be easily computed using the pre-existing methods in the code. 
Lattice-dynamical calculations were performed on a $[16,16,16]$ k-point mesh. In Fig.~\ref{fig:VDOS} we display the phonon dispersions of the material and the Vibrational Density of States (VDOS), computed with a Gaussian broadening function with a standard deviation of $0.3$ THz. The pink area surrounding the phonon dispersion represents twice the linewidths---computed at $250$ K on a coarse grid and Fourier-interpolated on a finer one---highlighting the regions in reciprocal space where the separation between phonon bands is comparable with the vibrational broadening, and interband contributions to the heat conductivity are expected to be important, see Eqs. (\ref{eq:WTE}-\ref{eq:I-QHGK}). }
% This gives a graphical interpretation to the interband term: when the distance between bands is within a few linewidths the interband contributions cannot be neglected. The dispersion is made with the angular frequency $\omega=2\pi\nu$ instead of the frequency since linewidths compare directly with $\omega$ and not $\nu$ in Eq.~\ref{eq:k_markovian}. 
%Computing the linewidths on a dense kpath such as the one in Fig.~\ref{fig:VDOS} is computationally expensive, therefore we used a standard technique of Fourier interpolation. The technique consists of Fourier transforming $\gamma_{\mathbf q}\to \gamma_{\mathbf R}$, where $\mathbf R$ is a vector of the Bravais lattice. Then the $\gamma_{\mathbf R}$ can be transformed back to $\gamma_{\mathbf q}$ with a denser k-point mesh or just along a kpath. In our case, the initial k-point mesh is the $[16,16,16]$ used for the computation of thermal conductivity.}

\bibliography{biblio}% Produces the bibliography via BibTeX.

\begin{thebibliography}{27}
\bibitem [{vMa()}]{vMatrixNote}%
  \BibitemOpen
  \href@noop {} {}\bibinfo {note} {The velocity matrix can always be assumed to
  be real, because so can be chosen the vibrational normal modes. When normal
  modes are degenerate, as it is the case, \emph{e.g.}, in periodic crystals,
  the matrix is not necessarily real, but can be made so by considering real
  linear combinations of vibrational modes with opposite crystal
  momentum.}\BibitemShut {Stop}%
\end{thebibliography}

\end{document}